%% file: main.tex
\documentclass[sigconf, nonacm]{acmart}

\usepackage{pifont} 
\usepackage{tikz}
\usetikzlibrary{decorations.pathreplacing,calc}

\AtBeginDocument{%
  }

\begin{document}

\title{Smart Fast Finish: Preventing Overdelivery via Daily Budget Pacing at DoorDash}


\author{Rohan Garg}
\email{rohan.garg@doordash.com}
\affiliation{%
  \institution{DoorDash, Purdue University}
  \city{San Francisco}
  \state{CA}
  \country{USA}
}

\author{Yongjin Xiao}
\email{yongjin.xiao@doordash.com}
\affiliation{%
  \institution{DoorDash}
  \city{San Francisco}
  \state{CA}
  \country{USA}
}

\author{Jason (Dianxia) Yang}
\email{jason.yang@doordash.com}
\affiliation{%
  \institution{DoorDash}
  \city{San Francisco}
  \state{CA}
  \country{USA}
}

\author{Mandar Rahurkar}
\email{mandar.rahurkar@doordash.com}
\affiliation{%
  \institution{DoorDash}
  \city{San Francisco}
  \state{CA}
  \country{USA}
}

\renewcommand{\shortauthors}{Garg et al.}

\begin{abstract}
    \input{abstract}
\end{abstract}
\maketitle

\input{intro}
\input{problem}
\input{SFF}
\input{challenges}

\input{results}
\input{conclusion}


\bibliographystyle{alpha}
\newcommand{\etalchar}[1]{$^{#1}$}

\end{document}

%% file: abstract.tex
We present a budget pacing feature called Smart Fast Finish (SFF). SFF builds upon the industry standard Fast Finish (FF) feature in budget pacing systems that depletes remaining advertising budget as quickly as possible towards the end of some fixed time period. SFF dynamically updates system parameters such as start time and throttle rate depending on historical ad-campaign data. SFF is currently in use at DoorDash, one of the largest delivery platforms in the US, and is part of its budget pacing system. We show via online budget-split experimentation data and offline simulations that SFF is a robust solution for overdelivery mitigation when pacing budget.

%% file: intro.tex
\section{Introduction}

Online advertising platforms aim to allocate advertising budgets efficiently, ensuring ads are displayed throughout the day. Pacing the use of the advertisers' budgets throughout the day, referred to as intraday pacing, requires carefully towing the line between underspend and overspend. 

A common technique to prevent underspend is to pace the budget throughout the day and \emph{if} there is remaining budget for the final few hours of the day, the platform tries to deplete that budget as quickly as possible. This rapid depletion is called ASAP Pacing and this portion of the day is called the Fast Finish (FF) period. Although the FF period can prevent underspend, it can introduce severe overspend as well. 

The natural question to ask is, ``if we prevent advertisers from participating in ad auctions when they have depleted their budget entirely, how can we run into overspend issues?''

The reason this occurs is due to \emph{delayed attribution}. Delayed attribution occurs when we display an ad for an advertiser prior to their budget being depleted but the click (and purchase) occurs after the budget has been depleted. Since the click or purchase occurred, we continue to bill the advertiser by removing the charge from their monthly budget as opposed to their already depleted daily budget. This overspend due to delayed attribution is called \emph{overdelivery}. If we overdeliver on a day-to-day time frame by a small margin, this can leave an ad campaign with no budget towards the end of a larger time frame, say a month. This leads to multiple days in a row at the end of the month where the advertiser is served no ads whatsoever, often referred to as \emph{blackout}.

To better improve advertiser experience and combat overdelivery due to delayed attribution, we introduce Smart Fast Finish (SFF) as an iteration to the FF budget pacing feature. Our system incorporates dynamic fast finish adjustments and a probabilistic throttling mechanism to improve budget pacing. We demonstrate through simulations and online experiments that this approach can substantially reduce overspend while maintaining high ad campaign visibility throughout the day. Our data is from the DoorDash advertising platform. DoorDash is one of the largest delivery platforms in the United States and its advertising platform hosts hundreds of thousands of ad campaigns daily.


Charge-per-conversion billing is less common than charge-per-click based advertising platforms and the overdelivery problem is most severe for charge-per-conversion platfors due to the delayed attribution problem that occurs between click and conversion. DoorDash continues to use the charge-per-conversion model as it better aligns incentives between the platform and the advertiser. Some notable downstream effects of this decision are that the platform does not become overrun with sponsored content and that the platform and advertiser have the same goal - to make a sale. 
\subsection{Related Work}
Budget pacing is a key component of online advertising systems and as a result has attracted interest from both practitioners and theoreticians alike. 

A seminal paper that describes budget pacing in an industry setting is~\cite{agarwal2014budget}. In ~\cite{agarwal2014budget}, the authors describe a budget pacing system implemented at LinkedIn and they introduce Fast Finish. Their implementation paces budget so as to deplete the budget in 22 hours. If there is remaining budget for the final 2 hours of the day, they remove all pacing and deplete budget as quickly as possible. More recently, teams at Yahoo and eBay describe new techniques and data for budget pacing systems~\cite{mystique2024,NguyenGBAB23}.
In ~\cite{mystique2024}, the authors describe a budget pacing system in place at Yahoo that uses a soft throttling approach that allows all ad requests to participate in the ad auction with a modified bid.
In ~\cite{NguyenGBAB23}, the authors give a test bed of sponsored search data from the eBay marketplace for evaluating different budget pacing techniques. In ~\cite{bidOpt2019}, the authors establish the theoretical foundation and performance guarantees of the PID controller in ad budget pacing. However, we highlight that ~\cite{bidOpt2019}  uses PID for bid value adjustment, not for controlling timing of budget pacing. Our approach shares similarities in using a cumulative error estimate to guide pacing (our approach uses historical overspend), but the methodologies are distinct. 

On the theoretical side, ~\cite{pmlr-v202-balseiro23a} study the sample complexity of learning a good budget expenditure plan and ~\cite{pmlr-v235-balseiro24a} study the varying degrees of coupling of the autobidder feature and the budget pacing feature of an ads delivery system. We also note some work done in autobidding. These papers solve similar problems of control and timing of bids, auctions, and charges~\cite{NEURIPS2024_autoVahab,NEURIPS2021_autoBalseiro,10.1145/3699824.3699838},.

The work in this paper also appears in a blog post published on the DoorDash Engineering Blog~\cite{doordash2024interns}.

\subsection{Our Contributions}

We present the Smart Fast Finish component of a budget pacing system that robustly reduces overdelivery of ad spend by dynamically adjusting pacing behavior towards the end of a pre-defined time window. The various advances made towards designing SFF are listed below:

\begin{enumerate}
    \item We give a smooth, non-linear function that takes the historical overspend average for a particular ad-campaign and outputs a value $t \in [0,1]$ which represents what time in the window to start SFF.
    \item We describe how to incorporate the dynamic start time with a transition window that slowly reduces the ad throttle rate.
    \item We describe engineering challenges faced and their solutions while integrating SFF into DoorDash's budget pacing system.
    \item We give offline simulation data and online budget-split experimentation data showing the effectiveness of SFF.
    \item We give open research and engineering directions to improve large-scale ads budget pacing systems.
\end{enumerate}

%% file: problem.tex
\section{Problem Definition}

In this work, we aim to minimize overdelivery that occurs due to delayed attribution without throttling spend so much that we see consistent underspend. 

\subsection{The Problem}
At DoorDash, when a consumer clicks on an advertisement, we charge that advertiser an ad fee. This fee is taken from the advertiser's daily budget. Because of daily fluctuations in user traffic, however, we also set our own daily spend goals that can vary from day to day. Once the daily goal is hit, we stop serving ads for that advertiser. We have some flexibility on daily advertiser budget spend on the most we can spend in a given day. This gap allows for some overspending, up to  a maximum amount called \emph{the billing cap}, even after the daily spend goal has been achieved. Overspend occurs when a consumer clicks on an ad before the daily spend goal has been reached, but then fails to place an order until after that goal is reached. i.e. delayed attribution. If too many orders contribute to this delayed attribution, we spend past our daily goal, resulting in ad overspend. This can be particularly severe for enterprise campaigns as they attract a large number of consumers and tend to have high conversion rates.

\begin{figure}[ht]
\centering
\includegraphics[width=1\columnwidth]{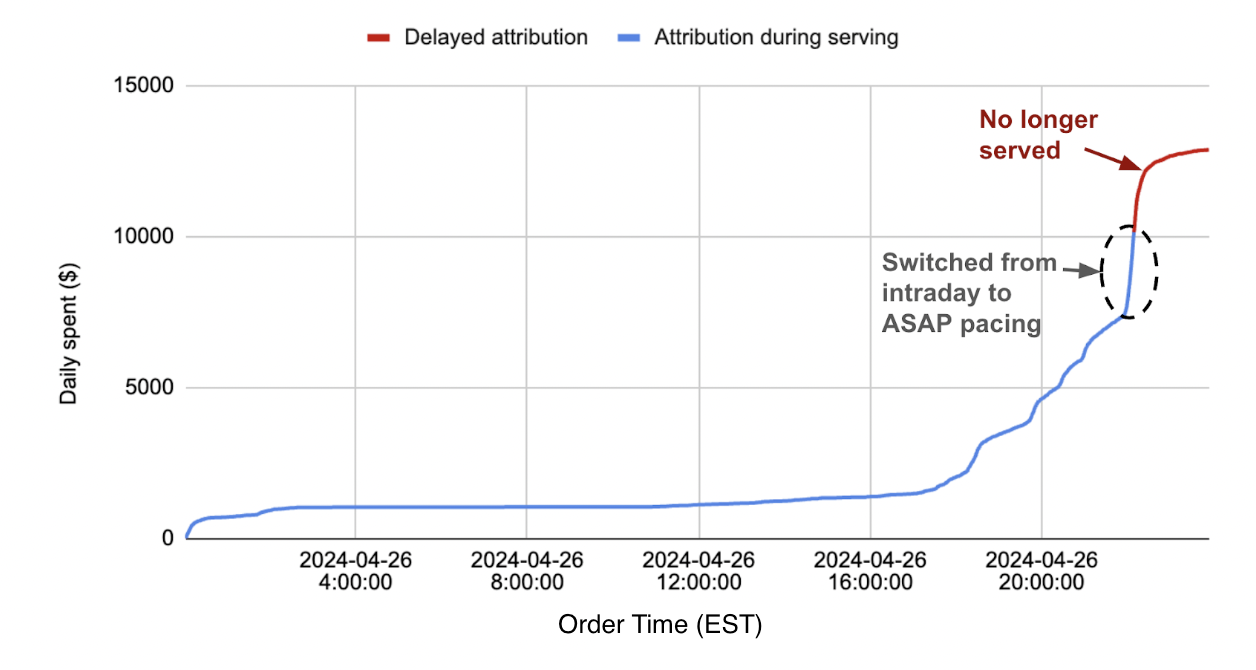} 
\caption{Daily spend curve for an example campaign from 2024. The blue line shows spend before reaching the daily goal while the red line shows spend due to delayed attribution, or over-delivery.}
\label{fig: overdelivery spend curve}
\end{figure}

Although we aim to minimize ad overspend, we must be careful not to throttle spending so much that we cause underspend. Our current method to prevent this is to switch from the intraday pacing system for most of the day to ASAP pacing for the last few hours of the day i.e. the Fast Finish period. The Fast Finish period can cause a sharp rise in ad spend, as shown in Figure~\ref{fig: overdelivery spend curve}. As a result, high-quality campaigns show a big spike in spending right after the FF period - a negative experience for our advertisers. Our goal in this project was to iterate on our intraday pacing system and the Fast Finish mechanism so that we can minimize overspend while avoiding underspend.

Improving budget pacing systems is a key challenge for DoorDash and computational advertising in general. Poor budget pacing can result in unstable impression counts, reduced revenue, and blackout days (where an ad campaign gets no ad impressions for many days at the end of the month). Our work makes progress towards a more robust and well-rounded budget pacing system.

\subsection{Business Opportunity}

To understand the business opportunity as well as the severity of advertiser overspend at DoorDash, we analyzed the ads platform data and specifically the data surrounding delayed attribution. For the last week of May 2024, enterprise ad campaigns experienced about 11.5\% overspending beyond goals for the week. Because budgets can be depleted earlier in the month, we studied spending data for the first week of June 2024 as well. The data showed that in the first week of June, enterprise campaigns saw nearly 25\% ad overspend. The data was clear - many advertisers on the DoorDash platform were struggling with severe ad overspend due to delayed attribution. 

%% file: SFF.tex
\section{Smart Fast Finish}
In this section we present the Smart Fast Finish (SFF) mechanism and its two primary components: dynamically setting the SFF start time and using a transition window to slowly reduce the throttle rate to zero percent. At a high level, we use the average historical campaign-specific ratio of the actual daily spend over the daily spend goal over a past period of time (e.g 6 months) to dynamically set the SFF start time and couple that with the transition window to slowly reduce the throttle rate.

\subsection{Setting SFF Start Time}

The first component of SFF dynamically sets the start time for when Smart Fast Finish begins. The standard FF implementation simply started at a set time if there was remaining budget at that time. This set time was the same across all ad campaigns. By dynamically adjusting the start time, we would set the start time later in the day for campaigns that historically were overspending and earlier in the day for campaigns that historically were underspending.

The rationale behind this was to allow our intraday pacing algorithm more time to serve high-spending campaigns. By allowing more time for the daily spend, we would likely hit the goal before we would need to trigger smart fast finish, leading to fewer delayed attributions.

To set the SFF start time $t$, we used the following function:

\begin{align}
    t = \textit{D} \cdot \sqrt{\textit{OR}-\textit{min-OR}} + \textit{min-start}
\end{align}
\label{eq: sff-start-time}

where \textit{OR} is the campaign's overspend ratio over a past time period (e.g the past 6 months) and $D$ is a constant defined as:

\begin{align}
    D = \frac{(\textit{max-start} - \textit{min-start})}{\sqrt{\textit{max-OR} - \textit{min-OR}}}
\end{align}

In these equations, \textit{min-OR} and \textit{max-OR} are the minimum and maximum campaign overspend ratios for which we want to apply a dynamic start time change to. For example, we may decide that if a campaign shows only $2\%$ overspend in the last month, we do not want to apply any change to their SFF start time. As a result, we set the \textit{min-OR} value to $1.03$. On the other hand, at a certain overspend ratio, we may decide that we want to give the intraday pacing system as much time as possible and delay the start of SFF as much as we can. In this case, we may desire that all campaigns that show more than, say, $50\%$ overspend should only start the SFF for the last $5\%$ of the day. In this case, we would set \textit{max-OR} to $1.50$.

Similar to \textit{min-OR} and \textit{max-OR}, we have \textit{min-start} and \textit{max-start} which define the earliest and latest \emph{start} times for the SFF start time. For example, if we wish that no campaign can begin SFF before $85\%$ of the day, we set \textit{min-start} to $0.85$. Similarly, if we wish that no campaign can begin SFF after $95\%$ of the day, we set \textit{max-start} to $0.95$.

\subsection{Throttle Rate Transition Window}

The second component of SFF changes the throttle rate of participation in ads auctions. Instead of immediately dropping the probability of participation in the ad auction to zero, we slowly drop it to zero over a set period of time, for example an hour. By slowing the drop of the throttle rate, we hypothesized that we could smooth the number of orders and ads spend as we approached the daily spend goal. We used a fixed percentage of campaign targeting hours as the window length and dropped the current throttle rate by a constant amount over each minute in the transition window down to zero by the end of the window. 

%% file: challenges.tex
\section{Challenges}
There were two primary technical challenges that occurred while implementing SFF at DoorDash.
\begin{enumerate}
    \item A positive feedback loop in our algorithm that would introduce unstable behavior.
    \item Embedding SFF into the budget pacing system (limited latency budget, caching, parallelization, etc.)
\end{enumerate}
\subsection{Feedback Loop}\label{feedback loop}
Our first challenge was that our algorithm was subject to a positive feedback loop that could cause unstable behavior. 

At a high level, we were using historical, campaign-specific, spend data to set our SFF start time that would then alter the spend data used by the algorithm in the future. To illustrate this, consider a campaign that has been overspending. When we dynamically adjust the SFF start time by starting the fast finish period \emph{later} in the day, we will reduce overspend. This in turn will cause the algorithm to think the campaign is not overspending as severely and will in turn adjust the SFF start time by setting the start time to be \emph{earlier} in the day. This cycle of unstable behavior would continue on and on. 

To combat this, we insisted that any updates to the start time only \emph{pushed the start time later in the day}. So, on a given day, if the SFF start time function wanted to move the start time \emph{earlier} in the day, this update would be ignored. This is coupled with completely re-calculating the start time once every pre-defined period of time, e.g once a week, once every 6 months etc. This choice of how often to refresh the SFF start time can vary depending on the application and the advertising platform. 

\subsection{Engineering Challenges}
The second issue involved nuances in engineering the algorithm and integrating it with our ad exchange service. To optimize the implementation, we utilized parallelization and caching wherever we could. Since we determine the dynamic FF start time once per day, we can calculate this using equation~\ref{eq: sff-start-time} once and cache this value for the day. This value, is retrieved, in parallel, along with other inputs required for intraday budget pacing so as not to consume too much latency budget.


%% file: results.tex
\section{Results}
In this section we cover the results of the offline simulations and online budget A/B experiment for Smart Fast Finish.

\subsection{Offline Simulations}

Our offline simulations helped us determine the details of the dynamic start time and transition window components of SFF. Our simulations are run on a single high-spending campaign from April 2023. On a high-level, the simulation took in the number of ad requests per minute and ran them through the pacing algorithm by throttling ad requests if we had hit the daily goal, and allowing them to proceed otherwise. If an ad request proceeds, it could take place in the auction and we would estimate the request winning the auction by using that ad's conversion rate and some added uniform noise. Finally, we plotted the spend curves which show the cumulative spend of the campaign throughout the day. By looking at the ASAP Pacing curve (no ad requests are throttled), we can see that the daily spend goal can be reached as quickly as within the first few hours of the day. Our goal of these simulations was to see the impact on campaign live hours, the total number of hours before the campaign hit its daily spend goal, that resulted from the various individual components of SFF. 

\begin{figure}[h]
\centering
\includegraphics[width=1\columnwidth]{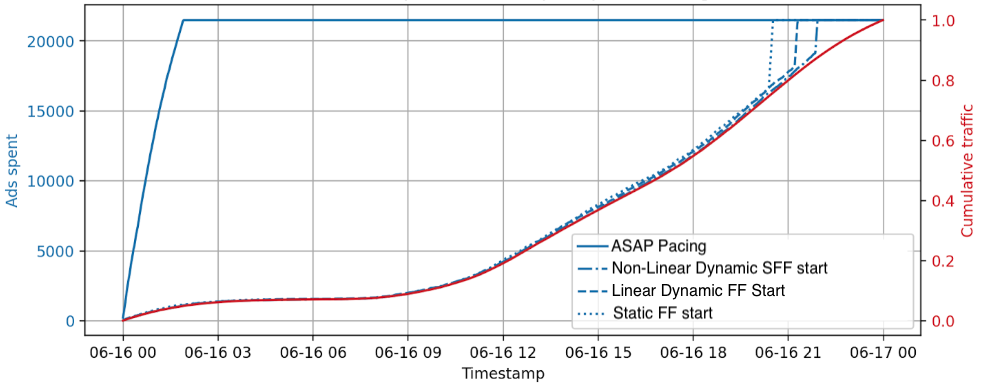}
\caption{Daily spend curve for an example campaign with and without dynamic FF start time.}
\label{fig: dynamic start spend curve}
\end{figure}

In figure~\ref{fig: dynamic start spend curve}, we first simulated the dynamic start time feature of SFF. We compared a function that was linear in the historical overspend ratio to the non-linear function presented in ~\ref{eq: sff-start-time}. The non-linear start-time function increased the campaign live hours of the campaign by 6\%.

\begin{figure}[h]
\centering
\includegraphics[width=1\columnwidth]{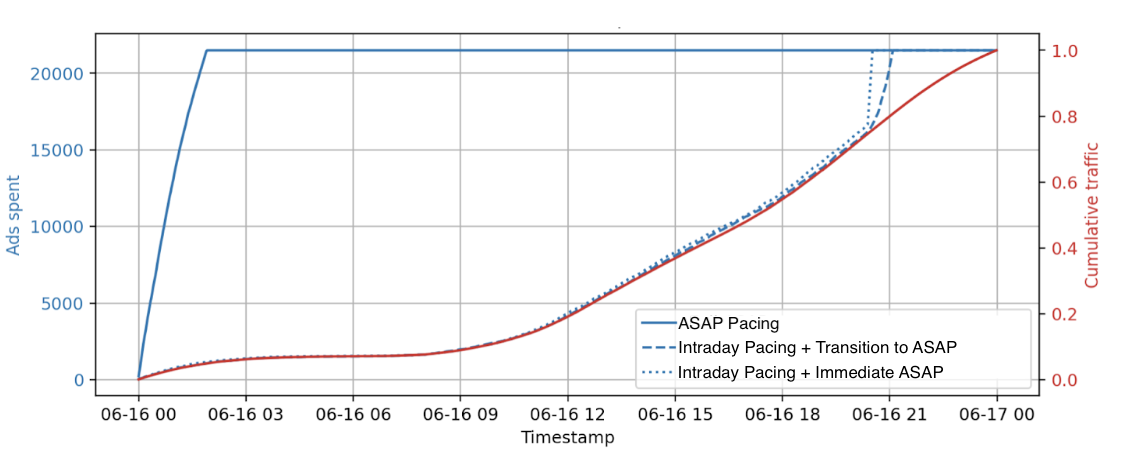} 
\caption{Daily spend curve for an example campaign with and without transition window for throttle rate drop.}
\label{fig: transition spend curve}
\end{figure}

In figure~\ref{fig: transition spend curve}, we simulated the transition window feature of SFF and compared it to running intraday pacing with an immediate drop to ASAP Pacing (equivalently, zero percent throttle rate). We saw a `smoother' spend curve to the daily goal as well as an increase in campaign live hours by 3\%. 

\begin{figure}[h]
\centering
\includegraphics[width=1\columnwidth]{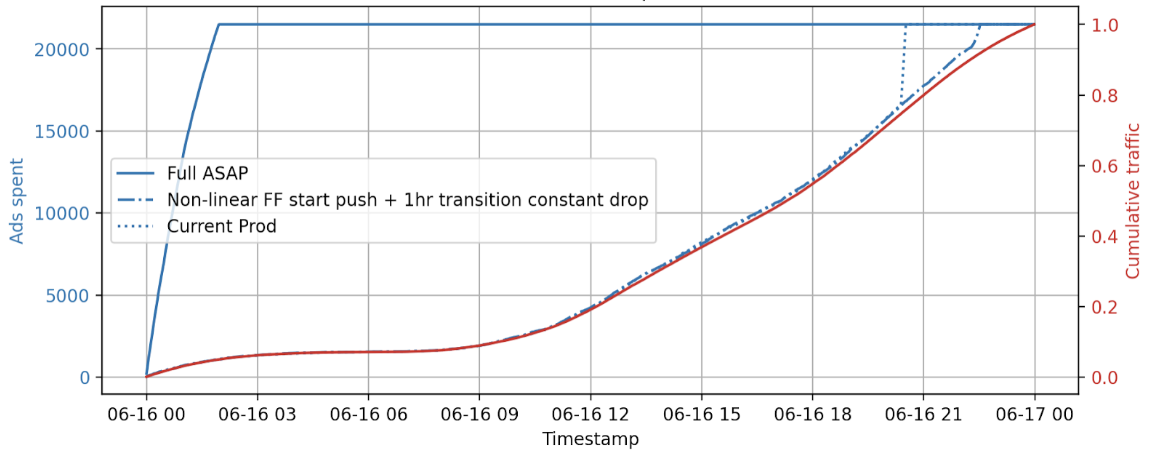} 
\caption{Daily spend curve for an example campaign with and without the entire SFF feature.}
\label{fig: SFF spend curve}
\end{figure}

Finally, in figure~\ref{fig: SFF spend curve}, we compare the entire SFF feature (dynamic start time component and the transition window component) to the system in production at the time of developing SFF. That production system was using intraday pacing with the traditional FF feature (static start time with immediate switch to ASAP Pacing). We saw an increase in campaign live hours by more than 8\% by using SFF.

Based on these simulations, we decided that Smart Fast Finish should be comprised of a non-linear function to dynamically set the start time and a 1 hour long transition window to slowly drop the throttle rate to zero percent.

\subsection{Online Experimentation}

After incorporating SFF into the budget pacing system, the new budget pacing system is now composed of three stages over the course of each day: intraday pacing, transition window, and fast finish.

\begin{figure}[htb]
\centering
\includegraphics[width=1\columnwidth]{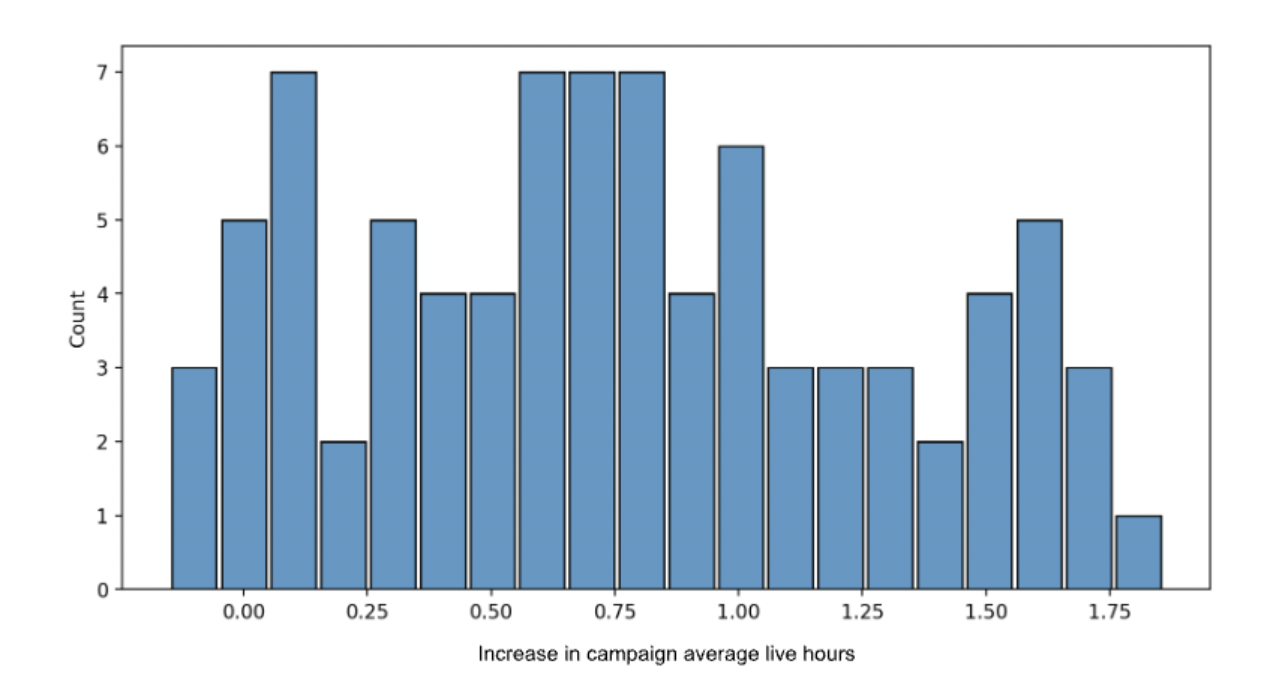} 
\caption{Histogram of increase in average campaign live hours. On average, we observed an increase by 45 min per day.}
\label{fig: camp-live-hours}
\end{figure}

\begin{figure}[htb]
\centering
\includegraphics[width=1\columnwidth]{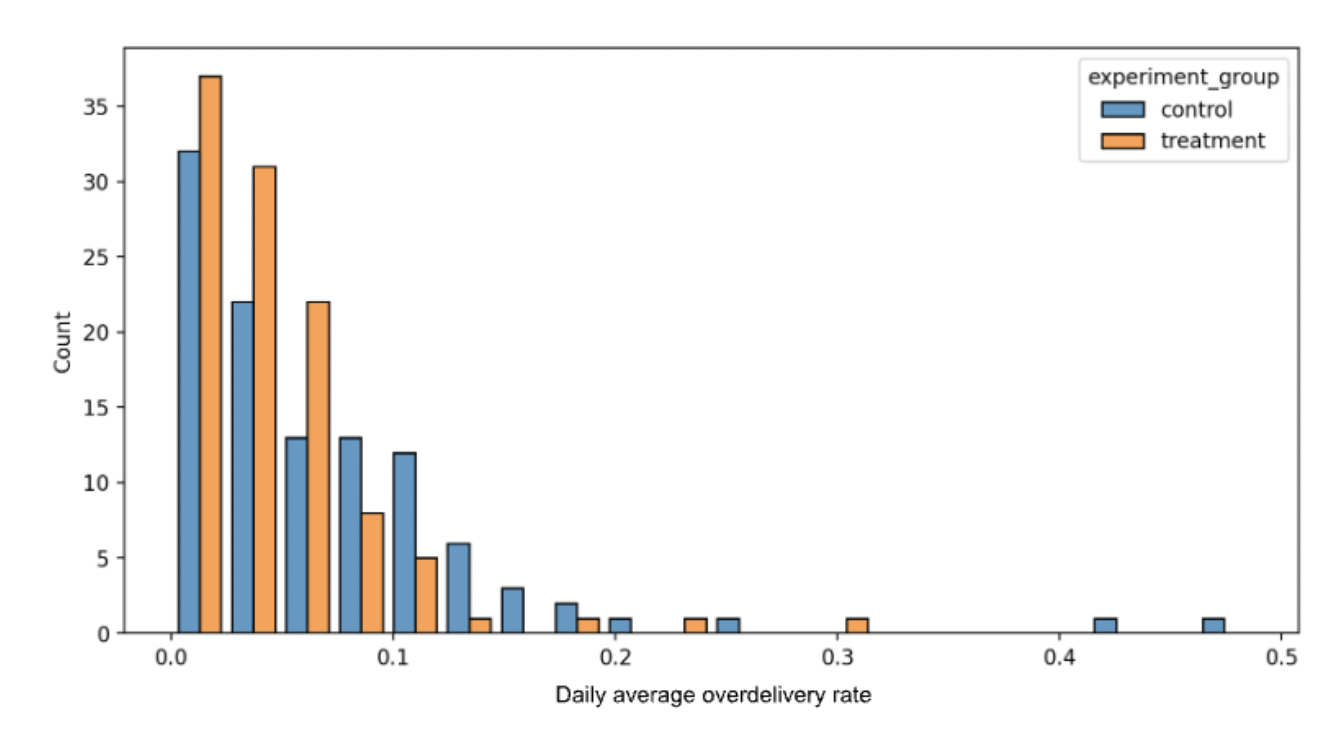} 
\caption{Histogram of showing average overdelivery rate. The treatment group (orange bars) are denser near 0, indicating less over-delivery.}
\label{fig: camp-overdelivery}
\end{figure}

\begin{figure}[htb]
\centering
\includegraphics[width=1\columnwidth]{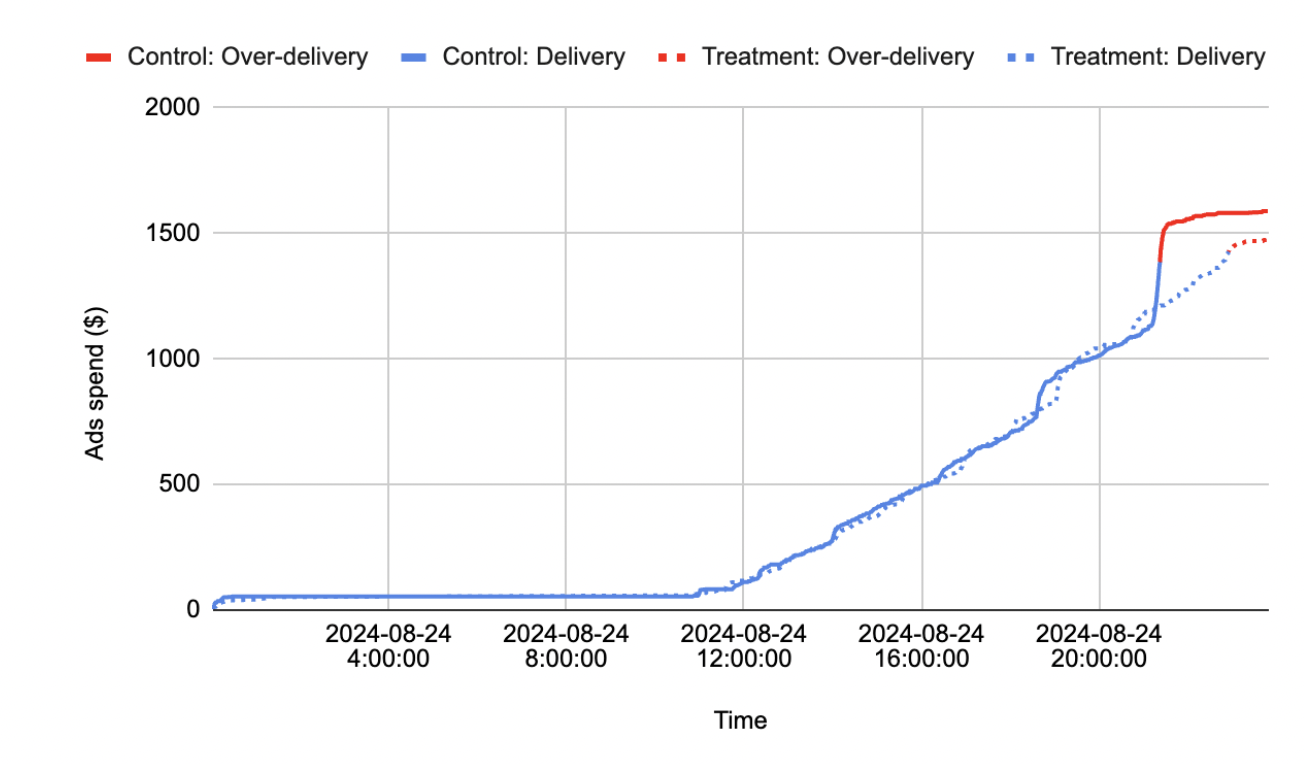} 
\caption{An example campaign daily spend curve. The treatment group showed a clear increase in campaign live hours (blue dots ended at 10:30 p.m., i.e. +1.1-hour later than the blue straight line) and also had less overdelivery, dropping from 10\% to 3\%. }
\label{fig: single-camp-spend}
\end{figure}

We conducted a one-week budget-split A/B experiment on \emph{all campaigns that are subject to intraday pacing} that demonstrated the results of our new algorithm, as shown in Figures \ref{fig: camp-live-hours}, \ref{fig: camp-overdelivery} , and \ref{fig: single-camp-spend}. Our budget-split experiment design follows that of~\cite{LiuKang2021BudgetSplitExp}. We took all campaigns that run through our budget pacing system and split their budget into two groups: the control group and the treatment group. The control group was subject to Intraday Pacing coupled with the traditional Fast Finish while the treatment group was subject to Intraday Pacing coupled with Smart Fast Finish.

In figure~\ref{fig: camp-live-hours}, we show the distribution in the increase in campaign live hours. The average increase in campaign live hours was 45 minutes, or 4.2\%. In figure~\ref{fig: camp-overdelivery}, we show the distributions of overdelivery of both the control and treatment groups. The average overdelivery rate was reduced from 7\% to 4.7\%, a decrease of 2.4\%. Finally, in figure~\ref{fig: single-camp-spend}, we show a single high-spending campaign that was part of the experiment. We see a clear increase of 1.1 hours in campaign live hours and a reduction in overdelivery from 10\% to 3\%.

We note here that since our experiments are on advertisers on our platform and not users, the sample size is constrained. As a result, it is not straightforward to get statistically significant results from the budget-split experimentation. However, we give convincing data that Smart Fast Finish is effective at mitigating overdelivery.

\begin{itemize}
  \item \textbf{All Campaigns:}
  \begin{itemize}
    \item Average campaign live hours increased by 45 minutes (+4.2\%)
    \item Average soft overdelivery dropped from 7.0\% to 4.7\% (–2.3\%)
  \end{itemize}
  \item \textbf{The Largest 28 campaigns:}
  \begin{itemize}
    \item Average total live hours increased by 48 minutes (+4.6\%)
    \item Overdelivery decreased from 7.7\% to 4.1\% (–3.6\%)
    \end{itemize}
\end{itemize}


%% file: conclusion.tex
\section{Conclusion}

Our results show that the data-driven Smart Fast Finish feature is a robust and effective solution to mitigating overdelivery in budget pacing systems. We presented the two components of SFF: the non-linear function that dynamically sets the start time and the transition window to reduce the throttle rate. Finally, we showed offline simulation and online budget A/B experiment results to verify our findings. We conclude with some open directions for further research:

\begin{enumerate}
    \item Can SFF be integrated into a budget pacing system that utilizes soft throttling instead of hard throttling~\cite{mystique2024}?
    \item Is there a solution to the feedback loop problem presented in Section~\ref{feedback loop} that does not require insisting on only increasing the SFF start time?
    \item Besides the historical overspend ratio, what other historical spend metrics could be used to improve SFF?
\end{enumerate}


%% file: main.bbl
\begin{thebibliography}{DMM{\etalchar{+}}24}

\bibitem[ABB{\etalchar{+}}24]{10.1145/3699824.3699838}
Gagan Aggarwal, Ashwinkumar Badanidiyuru, Santiago~R. Balseiro, Kshipra Bhawalkar, Yuan Deng, Zhe Feng, Gagan Goel, Christopher Liaw, Haihao Lu, Mohammad Mahdian, Jieming Mao, Aranyak Mehta, Vahab Mirrokni, Renato~Paes Leme, Andres Perlroth, Georgios Piliouras, Jon Schneider, Ariel Schvartzman, Balasubramanian Sivan, Kelly Spendlove, Yifeng Teng, Di~Wang, Hanrui Zhang, Mingfei Zhao, Wennan Zhu, and Song Zuo.
\newblock Auto-bidding and auctions in online advertising: A survey.
\newblock {\em SIGecom Exch.}, 22(1):159–183, October 2024.

\bibitem[AGWY14]{agarwal2014budget}
Deepak Agarwal, Souvik Ghosh, Kai Wei, and Siyu You.
\newblock Budget pacing for targeted online advertisements at linkedin.
\newblock In {\em Proceedings of the 20th ACM SIGKDD international conference on Knowledge discovery and data mining}, pages 1613--1619, 2014.

\bibitem[BBF{\etalchar{+}}24]{pmlr-v235-balseiro24a}
Santiago~R. Balseiro, Kshipra Bhawalkar, Zhe Feng, Haihao Lu, Vahab Mirrokni, Balasubramanian Sivan, and Di~Wang.
\newblock A field guide for pacing budget and {ROS} constraints.
\newblock In Ruslan Salakhutdinov, Zico Kolter, Katherine Heller, Adrian Weller, Nuria Oliver, Jonathan Scarlett, and Felix Berkenkamp, editors, {\em Proceedings of the 41st International Conference on Machine Learning}, volume 235 of {\em Proceedings of Machine Learning Research}, pages 2607--2638. PMLR, 21--27 Jul 2024.

\bibitem[BDM{\etalchar{+}}21]{NEURIPS2021_autoBalseiro}
Santiago Balseiro, Yuan Deng, Jieming Mao, Vahab Mirrokni, and Song Zuo.
\newblock Robust auction design in the auto-bidding world.
\newblock In M.~Ranzato, A.~Beygelzimer, Y.~Dauphin, P.S. Liang, and J.~Wortman Vaughan, editors, {\em Advances in Neural Information Processing Systems}, volume~34, pages 17777--17788. Curran Associates, Inc., 2021.

\bibitem[BKM{\etalchar{+}}23]{pmlr-v202-balseiro23a}
Santiago~R. Balseiro, Rachitesh Kumar, Vahab Mirrokni, Balasubramanian Sivan, and Di~Wang.
\newblock Robust budget pacing with a single sample.
\newblock In Andreas Krause, Emma Brunskill, Kyunghyun Cho, Barbara Engelhardt, Sivan Sabato, and Jonathan Scarlett, editors, {\em Proceedings of the 40th International Conference on Machine Learning}, volume 202 of {\em Proceedings of Machine Learning Research}, pages 1636--1659. PMLR, 23--29 Jul 2023.

\bibitem[DMM{\etalchar{+}}24]{NEURIPS2024_autoVahab}
Yuan Deng, Jieming Mao, Vahab Mirrokni, Hanrui Zhang, and Song Zuo.
\newblock Efficiency of the first-price auction in the autobidding world.
\newblock In A.~Globerson, L.~Mackey, D.~Belgrave, A.~Fan, U.~Paquet, J.~Tomczak, and C.~Zhang, editors, {\em Advances in Neural Information Processing Systems}, volume~37, pages 139270--139293. Curran Associates, Inc., 2024.

\bibitem[GNM{\etalchar{+}}24]{doordash2024interns}
Rohan Garg, Thao Nguyen, Anastasiya Masalava, Yufan~"Joy" Xiao, and Shanting Hou.
\newblock Part 1: Doordash 2024 summer intern projects, 2024.
\newblock Accessed: 2024-11-14.

\bibitem[LMK21]{LiuKang2021BudgetSplitExp}
Min Liu, Jialiang Mao, and Kang Kang.
\newblock Trustworthy and powerful online marketplace experimentation with budget-split design.
\newblock In {\em Proceedings of the 27th ACM SIGKDD Conference on Knowledge Discovery \& Data Mining}, KDD '21, page 3319–3329, New York, NY, USA, 2021. Association for Computing Machinery.

\bibitem[NGB{\etalchar{+}}23]{NguyenGBAB23}
Phuong~Ha Nguyen, Djordje Gligorijevic, Arnab Borah, Gajanan Adalinge, and Abraham Bagherjeiran.
\newblock Practical budget pacing algorithms and simulation test bed for ebay marketplace sponsored search.
\newblock In Abraham Bagherjeiran, Nemanja Djuric, Kuang{-}chih Lee, Linsey Pang, Vladan Radosavljevic, and Suju Rajan, editors, {\em Proceedings of the Workshop on Data Mining for Online Advertising (AdKDD 2023) co-located with the 29th {ACM} {SIGKDD} Conference on Knowledge Discovery and Data Mining {(KDD} 2023), Long Beach, CA, USA, August 7, 2023}, volume 3556 of {\em {CEUR} Workshop Proceedings}. CEUR-WS.org, 2023.

\bibitem[SAS{\etalchar{+}}24]{mystique2024}
Rotem Stram, Rani Abboud, Alex Shtoff, Oren Somekh, Ariel Raviv, and Yair Koren.
\newblock Mystique: A budget pacing system for performance optimization in online advertising.
\newblock In {\em Companion Proceedings of the ACM Web Conference 2024}, WWW '24, page 433–442, New York, NY, USA, 2024. Association for Computing Machinery.

\bibitem[YLW{\etalchar{+}}19]{bidOpt2019}
Xun Yang, Yasong Li, Hao Wang, Di~Wu, Qing Tan, Jian Xu, and Kun Gai.
\newblock Bid optimization by multivariable control in display advertising.
\newblock In {\em Proceedings of the 25th ACM SIGKDD International Conference on Knowledge Discovery \& Data Mining}, KDD '19, page 1966–1974, New York, NY, USA, 2019. Association for Computing Machinery.

\end{thebibliography}
